# Critical Success factors for Enterprise Resource Planning implementation in Indian Retail Industry: An Exploratory study

Poonam Garg
Professor, Information Technology and Management Dept.
Institute of Management Technology
Ghaziabad-India

*Abstract*— Enterprise resource Planning (ERP) has become a key business driver in today's world. Retailers are also trying to reap in the benefits of the ERP. In most large Indian Retail Industry ERP systems have replaced nonintegrated information systems with integrated and maintainable software. Retail ERP solution integrates demand and supply effectively to help improve bottom line. The implementation of ERP systems in such firms is a difficult task. So far, ERP implementations have yielded more failures than successes. Very few implementation failures are recorded in the literature because few companies wish to publicize their implementation failure. This paper explores and validates the existing literature empirically to find out the critical success factors that lead to the success of ERP in context to Indian retail industry. The findings of the results provide valuable insights for the researchers and practitioners who are interested in implementing Enterprise Resource Planning systems in retail industry, how best they can utilize their limited resources and to pay adequate attention to those factors that are most likely to have an impact upon the implementation of the ERP system.

## I. INTRODUCTION

An ERP system may be defined as a packaged business software system that enables a company to manage the efficient and effective use of resources (materials, human resources, finance, etc.) by providing an integrated solution for the organization's information processing needs (Nah *et al.*, 2001). ERP systems provide firms with two new and different types of functionality: a transaction processing function, allowing for the integrated management of data throughout the entire company, and a workflow management function controlling the numerous process flows within the company. ERP facilitates the flow of information between all the processes in an organization. ERP systems can also be an instrument for transforming functional organizations into process-oriented ones. When roperly integrated, ERP supports process-oriented businesses effectively (Al-Mashari, 2000). Recently, several practitioners have stated that ERP implementations have so far yielded more failures than successes in large organization.

Economic liberalization has brought about distinct changes in the life of urban people in India. A higher income group middle class is emerging in the Indian society. Demographic changes have also made palpable changes in social culture and lifestyle . In this environment Indian Retail Industry is witnessing rapid growth . AT Kearney has ranked India as fifth in terms of Retail attractiveness. Indian Retail Industry is the largest employer after Agriculture (around 8% of the population) and it has the highest outlet density in the world however this industry is still in a very nascent stage. The whole market is mostly unorganized and it is dominated by fragmented Kirana stores. A poor, supply chain and backward integration has weakened the whole process. A McKinsey report on India says organized retailing would increase the efficiency and productivity of entire gamut of economic activities, and would help in achieving higher GDP growth.

Enterprise resource Planning has become a key business driver in today's world. Retailers are also trying to reap in the benefits of the technology. Enterprise Resource Planning-ERP is, essentially, an integrated software solution used to manage a company's resources. Retailers are using ERP for product planning, parts purchasing, maintaining inventories, interacting with suppliers, providing customer service, and tracking orders. With ERP, retailers can save money in maintaining inventory, reduce the respondent time to the marketing demand, and get competence. More and more enterprises in the world are using it since its initial adoption.

A typical ERP implementation in a large retail industry takes between one and three years to complete and costs laks to crores. For these reasons, there is an urgent need to understand the underlying critical success factors that lead to the successful ERP implementations in such firms.

This paper is organized as follows. Section 2 describes the review of the literature on CSF of ERP implementation in context to retail industry. The third section and forth section of the paper describes the research objective and methodology adopted for this paper. The fifth section elaborates on the finding and describes the factors that play a role in success of ERP implementation in Retail industry. The last section draws some conclusion.

## II. CONSTRUCTS FROM LITERATURE REVIEW

Past studies have identified a variety of CSFs for ERP implementation, among which context related factors consistently appear. Some top CSFs which can be found frequently in literatures are including: well communicated top management commitment, "Best People" on team, "can-do" team attitude, other departments participation, clear goals and objectives, project management, reasonable expectations, project champion, vendor support, careful package selection, cooperation between enterprise and software company, User training and education, steering committee, strong vendor alliances, Effective communication, organizational size and structure, High Priority in company, Middle management commitment, Rapid, iterative prototyping to build knowledge, Initial " No modification" strategy, tight control on proposed modification, "seasoned", experienced consulting support, top management involvement etc. These factors have been found relevant as reported in some of the earlier studies.

In order to adopt a suitable research methodology, Following are the commonly identified CSFs identified by several researchers.

ERP teamwork and composition is important throughout the ERP life cycle. The ERP team should consist of the best people in the organization (Buckhout et al., 1999; Bingi et al., 1999; Rosario, 2000; Wee, 2000). Building a cross-functional team is also critical. The team should have a mix of consultants and internal staff so the internal staff can develop the necessary technical skills for design and implementation (Sumner, 1999). Both business and technical knowledge are essential for success (Bingi et al., 1999; Sumner, 1999). The ERP project should be their top and only priority (Wee, 2000). As far as possible, the team should be co-located together at an assigned location to facilitate working together (Wee, 2000). The team should be familiar with the business functions and products so they know what needs to be done to support major business processes (Rosario, 2000). The sharing of information within the company, particularly between the implementation partners, and between partnering companies is vital and requires partnership trust (Stefanou, 1999). Partnerships should be managed with regularly scheduled meetings. Incentives and risk-sharing agreements will aid in working together to achieve a similar goal (Wee, 2000). Top management support is needed throughout the implementation. The project must receive approval from top management (Bingi, 1999; Buckhout, 1999; Sumner, 1999) and align with strategic business goals (Sumner, 1999).
Top management needs to publicly and explicitly identify the project as a top priority (Wee, 2000). Senior management must be committed with its own involvement and willingness to allocate valuable resources to the implementation effort (Holland et al., 1999). Policies should be set by top management to establish new systems in the company. Business plan and vision Additionally, a clear business plan and vision to steer the direction of the project is needed throughout the ERP life cycle (Buckhout et al., 1999). There should be a clear business model of how the organization should operate behind the implementation effort (Holland et al., 1999 ), Project mission should be related to business needs and should be clearly stated (Roberts and Barrar, 1992). Goals and benefits should be identified and tracked (Holland et al., 1999). The business plan would make work easier and impact on work (Rosario, 2000). Effective communication is critical to ERP implementation (Falkowski et al., 1998). Expectations at every level need to be communicated. Management of communication, education and expectations are critical throughout the organization (Wee, 2000). Communication includes the formal promotion of project teams and the advertisement of project progress to the rest of the organization (Holland et al., 1999). Middle managers need to communicate its importance (Wee, 2000). Good project management is essential. An individual or group of people should be given responsibility to drive success in project management (Rosario, 2000). First, scope should be established (Rosario, 2000; Holland et al., 1999) and controlled (Rosario, 2000). The scope must be clearly defined and be limited. This includes the amount of the systems implemented, involvement of business units, and amount of business process reengineering needed. Any proposed changes should be evaluated against business benefits and, as far as possible, implemented at a later phase (Sumner, 1999; Wee, 2000). Additionally, scope expansion requests need to be assessed in terms of the additional time and cost of proposed changes (Sumner, 1999). Then the project must be formally defined in terms of its milestones (Holland et al., 1999). The critical paths of the project should be determined. Timeliness of project and the forcing of timely decisions should be managed (Rosario, 2000). Deadlines should be met to help stay within the schedule and budget and to maintain credibility (Wee, 2000). Project management should be disciplined with coordinated training and active human resource department involvement (Falkowski et al., 1998). Additionally, there should be planning of well-defined tasks and accurate estimation of required effort. The escalation of issues and conflicts should bemanaged (Rosario, 2000). Delivering early measures of success is important (Wee, 2000). Rapid ,successive and contained deliverables are critical. A focus on results and constant tracking of schedules and budgets against targets are also important (Wee, 2000). Employees should be told in advance the scope, objectives, activities and updates, and admit change will occur (Sumner, 1999). Project sponsor commitment is critical to drive consensus and to oversee the entire life cycle of implementation (Rosario, 2000). Someone should be placed in charge and the project leader should ``champion" the project throughout the organization (Sumner, 1999). There should be a high level executive sponsor who has the power to set goals and legitimize change (Falkowski et al., 1998). Sumner (1999) states that a business leader should be in charge so there is a business perspective. Transformational leadership is critical to success as well. The leader must continually strive to resolve conflicts and manage resistance. Change management is important, starting at the project phase and continuing throughout the entire life cycle. Enterprise wide culture and structure change should be managed (Falkowski et al., 1998), which include people, organization and culture change (Rosario, 2000). A culture with shared values and common aims is conducive to success. Organizations should have a strong corporate identity that is open to change. An emphasis on quality, a strong computing ability, and a strong willingness to accept new technology would aid in implementation efforts. Management should also have a strong commitment to use the system for achieving business aims (Roberts and Barrar, 1992). Users must be trained, and concerns must be addressed through regular communication, working with change agents, leveraging corporate culture and identifying job aids for different users (Rosario, 2000). As part of the change management efforts, users should be involved in design and implementation of business processes and the ERP system, and formal education and training should be provided to help them do so (Bingi et al., 1999; Holland et al., 1999). Education should be a priority from the beginning of the project, and money and time should be spent on various forms of education and training (Roberts and Barrar, 1992). Training, reskilling and professional development of the IT workforce is critical. User training should be emphasized, with heavy investment in training and reskilling of developers in software design and methodology (Sumner, 1999). Employees need training to understand how the system will change business processes. There should be extra training and on-site support for staff as well as managers during implementation. A support organization (e.g. help desk, online user manual) is also critical to meet users' needs after installation (Wee, 2000). Another important factor that begins at the project phase is BPR and minimum customization. It is inevitable that business processes are molded to fit the new system (Bingi et al., 1999). Aligning the business process to the software implementation is critical (Holland et al., 1999; Sumner, 1999). Organizations should be willing to change the business to fit the software with minimal customization (Holland et al., 1999; Roberts and Barrar, 1992). Software should not be modified, as far as possible (Sumner, 1999). Modifications should be avoided to reduce errors and to take advantage of newer versions and releases (Rosario, 2000). Process modeling tools help aid customizing business processes without changing software code (Holland et al., 1999). Broad reengineering should begin before choosing a system. In conjunction with configuration, a large amount of reengineering should take place iteratively to take advantage of improvements from the new system. Then when the system is in use reengineering should be carried out with new ideas (Wee, 2000). Quality of business process review and redesign is important (Rosario,2000). In choosing the package, vendor support and the number of previous implementers should be taken into account (Roberts and Barrar, 1992). Software development, testing and troubleshooting is essential, beginning in the project phase. The overall ERP architecture should be established before deployment, taking into account the most important requirements of the implementation. This prevents reconfiguration at every stage of implementation (Wee, 2000). There is a choice to be made on the level of functionality and approach to link the system to legacy systems. In addition, to best meet business needs, companies may integrate other specialized software products with the ERP suite. Interfaces for commercial software applications or legacy systems may need to be developed in-house if they are not available in the market (Bingi et al., 1999). Troubleshooting errors is critical (Holland et al., 1999). The organization implementing ERP should work well with vendors and consultants to resolve software problems. Quick response, patience, perseverance, problem solving and firefighting capabilities are important (Rosario, 2000). Vigorous and sophisticated software testing eases implementation (Rosario, 2000). (Scheer and Habermann, 2000) indicate that modeling methods, architecture and tools are critical. Requirements definition can be created and system requirements definition can be documented. There should be a plan for migrating and cleaning up data (Rosario, 2000). Proper tools and techniques and skill to use those tools will aid in ERP success (Rosario, 2000).

### III. OBJECTIVE OF THE STUDY

The objective of this paper is identify and validate the critical success factors empirically for ensuring successful implementation of Enterprise Resource Planning (ERP) packages in context to retail industry in India.

### IV. RESEARCH METHODOLOGY

The research process involved the following steps. First, a literature review was undertaken to identify what parameters to consider in research. It outlines the previous research and critical success factors for ERP implementation in retail industry were studied. Second, questionnaire was constructed and it was piloted. Last in depth interviews were held (with firm which have implemented ERP) to establish the evaluation criteria and factors were identified which result in Critical factors for ERP implementation in retail industry.

Reviewing the existing literature in ERP, we find out that 51 success factors have been recognized and studied. Further investigation revealed that 22 success factors were more frequently mentioned and studied in the previous research. The questionnaire which was developed for this research was based on these 22 CSF and the scaled used was a 5 Level Likert Scale. To ensure data validity and reliability of the survey instrument, an iterative process of personal interview with eight knowledgeable individuals (i.e. two IS faculty, two ERP supplier, two ERP consultant and two managerial level user) were conducted to modify the questionnaire before sending it out and their comments also helped us improve its quality. The questionnaires were sent to the ERP project managers and senior project team members of selected companies.

In this study, only organizations with prior experience of implementing ERP systems were selected as our investigative sample. The questionnaire was administered on 355 respondents out of which 110 questionnaires were completed, in which respondents were asked to indicate their level of importance for each of the construct items (critical success factors) using their response on a five point scale. The raw data was captured in a spread sheet software package. The spread sheet was then transported to software statistical package (SPSS).

Exploratory factor analysis (EFA) was used to summarize the 22 variables into smaller sets of linear composites that preserved most of the information in the original data set. A five factor solution best described the data. The resulting five factors namely, Top management, product selection, project management, team composition, training & education are shown in **Table I.** The component co variance matrix further shows that the three factors are not related to each further confirming the results of factor analysis **Table II.**

TABLE I. RESULTS OF EXPLORATORY FACTOR ANALYSIS

| Factor 1<br>Top Mangement | Factor Score | Factor 2<br>Product selection | Factor Score | Factor 3<br>Project Management | Factor Score | Factor 4<br>Team Composition | Factor score | Factor 5<br>Training & education | Factor score |
|---|---|---|---|---|---|---|---|---|---|
| Top mgmt commitment | .953 | Vendor support for implementation | .666 | Clear goal and objective | .896 | Can do attitude | .835 | User involvement | .714 |
| Steering commitee | .875 | Appropriate selection of ERP package | .726 | Effective project mgmt | .933 | Bright people | .707 | Education & training | .805 |
| Project champion | .904 | Package is user friendly | .777 | Reasonable expectation | .895 | | | Change management | .866 |
| High priority in company | .872 | Adequate scalability features | .746 | Other dept. participation | .837 | | | | |
| | | Organization size and structure | .745 | Change request | .912 | | | | |
| | | Suitability of H/W | .639 | Implementation strategy | .953 | | | | |
| | | | | Data conversion | .734 | | | | |
| | | | | Clear & effective communication | .937 | | | | |

TABLE II. COMPONENT SCORE COVARIANCE MATRIX

| Component | 1 | 2 | 3 | 4 | 5 |
|---|---|---|---|---|---|
| 1 | 1.000 | .000 | .000 | .000 | .000 |
| 2 | .000 | 1.000 | .000 | .000 | .000 |
| 3 | .000 | .000 | 1.000 | .000 | .000 |
| 4 | .000 | .000 | .000 | 1.000 | .000 |
| 5 | .000 | .000 | .000 | .000 | 1.000 |

Extraction Method: Principal Component Analysis.   Rotation Method: Varimax with Kaiser Normalization.   Component Scores.

After five factors (dimensions) were extracted from conducting the EFA procedure, we interpreted the results by assigning labels to the factors. The underlying factors were labeled as follows:

- Factor 1- Top management commitment: This includes 4 items that deal with the importance of top management support in the implementation of ERP in retail industry.
- Factor 2- Product selection: This consists of 6 items that relate to various selection criteria for ERP product.
- Factor 3- Project Management: This consists of 8 items that are very important for successful ERP implementation in Retail industry.
- Factor 4- Team composition: This is comprised of 2 items that deal with the importance of team in ERP implementation.
- Factor 5- Training and education: This includes 3 items that relate to the training and change management.

V. RESULTS AND IMPLICATIONS FOR RETAIL INDUSTRY

This study has identified the critical success factor (CSF) of ERP implementation in retail sector of India. These CSFs are classified into the following five dimensions: Top management, product selection, project management, team composition, training & education. Each dimension is described as follows:

*A. Top management*

The commitment of top management has been recognized as one of the most important elements in the successful implementation of ERP systems. Since the primary responsibility of top management is to provide sufficient financial support and adequate resources for building a successful system, the support of management will ensure that the project has a high priority within the organization and that it will receive the required resources and attention. The lack of financial support and adequate resources will inevitably lead to failure. Apart from this primary support, there should be steering committee, which can sponsor the money, ensure visibility and motivate the team.

*B. Product selection*

Implementation planning began with product selection. For successful ERP implementation, the retail industry must conduct a careful preliminary analysis and develop a plan for selecting the right ERP product for their organization. Implementing an ERP package is a complex and costly undertaking, so it's essential to choose the appropriate vendor, adequate scalability features, suitability of H/W and user friendliness of product depending on the size and structure of an organization.

*C. Project management*

A clear business vision is needed to guide the project throughout the ERP life cycle. Project management related factors like Clear goal and objective, Effective project management, Reasonable expectation, Other dept. participation, Change request, Implementation strategy, Data conversion, Clear & effective communication are very critical for a successful ERP implementation.

*D. Team composition*

Team composition includes the best and the brightest individuals from each functional area of the company. These individuals should understand the inner workings of their respective departments thoroughly. And the team must have can do attitude.

*E. Training and education*

Training and education are important for the successful implementation of any new system. Adequate training of the employees in an organization is important in allowing the benefits and advantages of using the ERP to be fully realized. In order to successful implementation any ERP system; a retail industry must establish a good fully functional change management. Change management are required to prepare the existing business's human resources and infrastructure to match ERP system requirement.

## VI. CONCLUSION

This study is valuable to researchers and practitioners interested in implementing Enterprise Resource Planning systems in retail industry. The EFA provides very interesting results by identifying the factors that actually have an impact on the successful implementation of ERP in retail industry. The findings from EFA identify items of importance that should help practitioners in their effort to implement ERP in retail industry.